\documentclass[preprint,amsmath,amssymb,aps,prd,showpacs,nofootinbib]{revtex4}

\newcommand{{\SD}}{\rm SD}

\newcommand{\vex}{\mbox{\boldmath${\rm x}$}}

\newcommand{\vep}{\mbox{\boldmath${\rm p}$}}

\newcommand{\veS}{\mbox{\boldmath${\rm S}$}}
\newcommand{\veL}{\mbox{\boldmath${\rm L}$}}

\newcommand{\vetau}{\mbox{\boldmath${\rm \tau}$}}

\newcommand{\veal}{\mbox{\boldmath${\rm \alpha}$}}

\newcommand{\lan}{\langle}
\newcommand{\ran}{\rangle}
\newcommand{\vephi}{\mbox{\boldmath${\rm \phi}$}}

\usepackage[dvips]{graphicx}
\usepackage[dvips]{color}

\newcommand{\rr}{\mathbf{r}}
\newcommand{\pp}{\mathbf{p}}

\newcommand{\xxi}{\mathcal{F}}

\begin{document}

\date{December 22, 2007}

\title{The chiral transitions in heavy-light mesons}

\author{A.M.Badalian}
\email{badalian@itep.ru}
\author{Yu.A.Simonov}
\email{simonov@itep.ru}
\author{M.A.Trusov}
\email{trusov@itep.ru}

\affiliation{ITEP, Moscow, Russia}


\begin{abstract}The mass shifts of the $P$-wave $D_s$ and $B_s$ mesons due
to coupling to $DK$,  $D^*K$ and $BK$, $B^*K$  channels are
studied using the chiral quark-pion Lagrangian without fitting
parameters. The strong  mass shifts down $\sim 140$ MeV and $\sim
100$ MeV for $D^*_s(0^+)$ and $D_s(1^{+'})$ and $\sim 100$ MeV for
$B^*_s(0^+)$ and $B_s(1^{+'})$ are calculated. Two factors are
essential for large mass shifts: strong coupling of the $0^+$ and
$1^{+'}$ states to the $S$-wave decay channel, containing a
Nambu-Goldstone meson, and the chiral flip transitions due to  the
bispinor structure of both heavy-light mesons. The masses
$M(B^*_s(0^+))=5695(10)$ MeV and $M(B_s(1^{+'}))=5730(15)$
MeV,very close to  $M(B(0^+))$ and $M(B(1^{+'}))$, are predicted.
Experimental limit on the width $\Gamma(D_{s1}(2536))<2.3$ MeV
puts strong restrictions on admittable mixing angle between the
$1^+$ and $1^{+'}$ states, $|\phi|<6^{\circ}$, which corresponds
to the mixing angle $\theta$ between the $^3P_1$ and $^1P_1$
states, $29^{\circ}<\theta< 41^{\circ}$.
\end{abstract}

\maketitle

\section{Introduction}
The heavy-light (HL) mesons  play a special role in hadron
spectroscopy. First of all, a HL meson is the simplest system,
containing one light quark in the field of almost static heavy
antiquark, and that allows to study quark (meson) chiral
properties. The discovery of the $D_s(2317)$ and $D_s(2460)$
mesons \cite{1,2} with surprisingly small widths and low  masses
has given an important impetus to study chiral dynamics and raised
the question why their masses are considerably lower than
expected values in different approaches: in relativistic quark
model calculations \cite{3I}--\cite{6I},  on the lattice
\cite{7I}, in QCD Sum Rules \cite{8I,9I}, in chiral models
\cite{10I}--\cite{12I}( for reviews see also \cite{13I,14I}). The
masses of $D_s(0^+)$ and $D_s(1^{+'})$ in closed-channel
approximation typically exceed by $\sim$ 140 and 90 MeV their experimental
numbers.

Thus  main theoretical goal  is  to understand dynamical mechanism
responsible for such large mass shifts of the $0^+$ and
$1^{+'}$ levels (both states have  the light quark orbital angular
momentum $l=0$ and $j=1/2$) and   explain why the position of
 other two levels (with $j=3/2)$ remains practically unchanged.
The importance of second fact has been underlined by S.Godfrey in
\cite{5I}.

 The mass shifts of the $D_s(0^+,1^{+'})$ mesons have already been
considered in a number of papers with the use of unitarized
coupled-channel model \cite{15I}, in nonrelativistic Cornell
model \cite{16I}, and in different chiral models
\cite{17I}--\cite{19I}. Here we address again this problem with the aim to
calculate also the mass  shifts of the $D_s(1^{+'})$ and
$B_s(0^+,1^{+'})$ states and the widths of the $2^+$ and $1^+$
states,  following the approach developed in \cite{18I}, for which
strong coupling to the S-wave decay channel, containing a pseudoscalar
($P$) Nambu-Goldstone (NG) meson, is crucially important.
Therefore in this approach principal difference exists between
vector-vector ($VV$) and $VP$ (or $PP$) channels. This analysis of
two-channel system is performed with the use of the
chiral quark-pion Lagrangian which has been  derived directly from the QCD
Lagrangian \cite{20I} and   does not contain fitting parameters,
so that the shift of the $D^*_s(0^+)$ state $\sim$ 140 MeV is
only determined by the conventional decay constant $f_K$.

Here the term "chiral dynamics" implies the mechanism by which in
the transition from one HL meson to another the octet of the NG
mesons $\phi$ is emitted. The corresponding Lagrangian $\Delta
L_{FGM}$,
\begin{equation}\Delta L_{FCM}=\bar q (\sigma r ) \exp (i\gamma_5
\phi/f_\pi)q,\label{1.1}\end{equation}contains the light-quark
part $\exp (i\gamma_5 \phi/f_\pi)$, where $\phi$ is the $SU(3)$
octet of NG mesons and the important factor $\gamma_5$ is present.
In the lowest order in $\phi$ this Lagrangian coincides with
well-known effective Lagrangian $\Delta L_{eff}$ suggested in
\cite{21I},\cite{22I}, where, however, an arbitrary constant $g_A$
is introduced . At large $N_c$, as argued in \cite{21I}, this
constant has to be equal unity, $g_A=1$. In \cite{10I,17I,22I}
this effective Lagrangian was applied to describe decays of HL
mesons taking $g_A<0.80$.

More general Lagrangian $\Delta L_{FCM}$ (\ref {1.1}) was derived
in the framework of the field  correlator method (FCM) \cite{20I,
23I}, in which the constant $g_A=1$ in all cases, and which
contains NG mesons to all orders, as seen from its explicit
expression (\ref{1.1}).

In Appendix A  with the use of the Dirac equation we show that
in the lowest order in $\phi$   $\Delta
L_{FCM}=\Delta L_{eff}$, if  indeed $g_A=1$. In our calculations
we always use $\Delta L_{FCM}$ with the $g_A= 1$ and derive the nonlinear
equation for the energy shift  and width, $\Delta E=\Delta\bar
E-\frac{\Gamma}{2}$, as in \cite{18I}. We do not assume any chiral
dynamics for the unperturbed levels, which are calculated here
with the use of the QCD string Hamiltonian \cite{24I,25I}, because
the mass  shift $\Delta E$ appears to be weakly dependent on the
position of unperturbed level.

It is essential that resulting shifts of the $J^P(0^+,1^{+'})$
levels are large only for the $D_s,B_s$ mesons, which lie close to
the $DK,D^*K,BK,B^*K$ thresholds, but not for the $D(1P),B(1P)$
mesons, in this way violating symmetry between them (this symmetry
is possible in close-channel approximation). In our calculations
 shifted  masses of the $D_s(0^+)$  and
$B_s(0^+)$ practically coincide  with those for the $D(0^+)$ and
$B(0^+)$, in agreement with the experimental fact that
$M_{\exp}(D(0^+))= 2350\pm 50$ MeV \cite{26I} is equal or even
larger than $M_{\exp}(D_s(0^+))=2317$ MeV. The states with
$j=3/2~~ D_s(1^+,2^+)$ and $D(1^+,2^+)$ have no mass
shifts and for them the mass difference is $\sim 100$ MeV, that
just corresponds to the mass difference between the $s$ and
light quark dynamical masses.

For the $D_s(1^{+'})$ and  $B_s(1^{+'})$  mesons  calculated
masses are also close to those of the  $D$ and $B$ mesons.
Therefore for given  chiral
dynamics the $J^P(0^+,1^{+'})$ states cannot be considered as the
chiral partners of the ground-state multiplet $J^P(0^-,1^-)$, as
suggested in \cite{11I}.

We also analyse why two other members of the 1P multiplet, with
$J^P=2^+$ and $1^+$, do not acquire the mass shifts due to decay
channel coupling (DCC) and have small widths. Such situation
occurs if the states $1^+$ and $1^{+'}$ appear to be almost pure
$j=\frac32$ and $j=\frac12$ states. Still small mixing angle
between them, $|\phi|<6^{\circ}$, is shown to be compatible with
experimental restriction on the width of $D_{s1}(2536)$, admitting
possible admixture of other component in the wave function (w.f.)
$\lesssim 10\%$.

In our analysis  the 4-component (Dirac) structure of the light
quark w.f. is crucially important. Specifically, the
emission of a NG meson is accompanied with the $\gamma_5$ factor
which permutes higher and lower components of the Dirac
bispinors. For the $j=1/2,P$ -wave and the $j=1/2,S$ -wave states it is
exactly the case that this "permuted overlap" of the w.f.
is maximal because the lower component of the first
state is similar to the higher component of the second state and
vice versa. We do not know other examples of such a "fine
tuning".On the other hand in the first approximation we neglect an
interaction between two mesons in the continuum, like $DK$,etc.

In present paper we concentrate on the $P$-wave $B,B_s$ mesons and
the effects of the  channel coupling. While the 1P levels of  the
$D, D_s$ mesons are now established with  good accuracy
\cite{1},\cite{2},\cite{26I}, for the  $B, B_s$  mesons only
relatively narrow $2^+,1^+$ states  have been recently observed
\cite{27I},\cite{28I}. According to these data the splitting
between the $2^+$ and $1^+$ levels is small, $\sim 20-10$ MeV,
while the mass
difference between $B_s(2^+)$ and $B(2^+)$ states is again $\sim
100$ MeV, as for the $D_s(2^+)$ and $D(2^+)$ mesons.

The actual position of the $B(1P),B_s(1P)$ levels is  important
for several reasons. Firstly, since dynamics  of $(q\bar b)$
mesons is very similar to that of $q\bar c$, the observation of
predicted large mass shifts of  the $B_s(0^+,1^{+'})$ levels  would
give a strong argument in favour of  the decay channel mechanism
suggested here and in \cite{18I}. Secondly, observation of all
$P$-wave states for the $B$,$B_s$ mesons could  clarify many
unclear features  of spin-orbit and tensor interactions in mesons.
Understanding of  the decay channel coupling (DCC) mass shifts could
become an important step in constructing  chiral theory of strong
decays with emission of one or several NG particles.

The paper is organized as follows. In the next Section we discuss
the formalism from \cite{18I}, extending that  to the case of the
$B,$ and $B_s$ mesons  and also to the $1^+$ states, and discuss
the mixing between the $1^+$ and $1^{+'}$ states. In
Section 3 the masses of HL mesons, calculated  in closed-channel
approximation, are  given. The Section 4
is devoted to the mechanism of chiral transitions while in Section
5 our calculations of the mass shifts due to DCC are
presented.The predictions of  the $B(J^P), B_s(J^P)$ masses
and discussion of our results are given in Section 6, while
Section 7 contains the Conclusions. In Appendix A a  connection between
the lowest order of $\Delta L_{FCM}$ and the effective Lagrangian
is illustrated. In Appendix B the details of our calculations of the
masses are given, while in Appendix  C the connection between FS
splittings and the mixing matrix (and angle) of the $1^+$ states is
discussed.

\section{Mixing of the $1^+$ and $1^{+'}$ states}

It is well known that  in single-channel approximation,
due to spin-orbit and tensor interactions the $P$-wave multiplet of a HL
meson is splitted into four levels with $J^P =0^+, 1^+_L, 1^+_H, 2^+$
\cite{29I}. Here for the $1^+$ states we use the notation H(L)
for the higher(lower) eigenstate of the mixing matrix
because apriori one cannot say which of them mostly consists of the light
quark $j=1/2$ contribution (see Appendix C). For  a HL meson,
strongly coupled to a nearby decay channel (DC),some member(s) of the
$P$-wave multiplet can be shifted down while another not.Just
such situationis takes place for the $D_s(1P)$ multiplet.

The scheme of classification, more adopted to a HL meson, in the
first approximation treats the heavy quark as a static one and
therefore the Dirac equation can be used to define
the light quark levels and wave functions \cite{10I}. Starting with the
Dirac's $P$-wave levels, one has the states with $j=1/2$ and $j=3/2$.
Since the light quark momentum $j$ and the quantum number $\varkappa$ are
conserved,\footnote{we use here the standard notation $ \varkappa =\mp
|j+\frac12|$ for $ j=\left\{\begin{array}{l} l+
\frac12\\l-\frac12\end{array}\right.$} they
run along the following possible values:

\begin{equation}
\begin{array}{ccc}
\begin{tabular}{c|c|c|c} \multicolumn{4}{c}{$D$}
\\ \hline $J^P$  & $j$ & $l$ & $\varkappa$ \\ \hline \hline $0^-$
 & $\frac{1}{2}$ & 0 & -1 \\ \hline $1^-$
 & $\frac{1}{2}$ & 0 & -1 \\ \hline \hline $1^-$
 & $\frac{3}{2}$ & 2 & +2
\end{tabular} & \hphantom{aaaaaa} &
\begin{tabular}{c|c|c|c} \multicolumn{4}{c}{$D_s$}
\\ \hline $J^P$  & $j$ & $l$ & $\varkappa$ \\
\hline \hline $0^+$  & $\frac{1}{2}$ & 1
& +1 \\ \hline $1^+$  & $\frac{1}{2}$ & 1 & +1 \\
\hline $1^+$  &
$\frac{3}{2}$ & 1 & -2 \\ \hline $2^+$  & $\frac{3}{2}$ & 1 & -2 \\
\hline \hline $2^+$  & $\frac{5}{2}$ & 3 & +3
\end{tabular}
\end{array}
\label{misha_table_3}
\end{equation}

The HL meson w.f. can be expressed in terms of the light quark
w.f. -- the Dirac bispinors $\psi^{jlM}_{q,s}$:
\begin{equation}
\Psi_D\left(J_{1/2}^-,M_f\right)=C^{J,M_f}_{\frac{1}{2},M_f-\frac{1}{2};\frac{1}{2},+\frac{1}{2}}
\psi_q^{\frac{1}{2},0,M_f-\frac{1}{2}}\otimes\bigl|\bar
c\uparrow\bigr\rangle+C^{J,M_f}_{\frac{1}{2},M_f+\frac{1}{2};\frac{1}{2},-\frac{1}{2}}
\psi_q^{\frac{1}{2},0,M_f+\frac{1}{2}}\otimes\bigl|\bar
c\downarrow\bigr\rangle, \end{equation}
\begin{equation}
\Psi_{D_s}\left(J_j^+,M_i\right)=C^{J,M_i}_{j,M_i-\frac{1}{2};\frac{1}{2},+\frac{1}{2}}
\psi_s^{j,1,M_i-\frac{1}{2}}\otimes\bigl|\bar
c\uparrow\bigr\rangle+C^{J,M_i}_{j,M_i+\frac{1}{2};\frac{1}{2},-\frac{1}{2}}
\psi_s^{j,1,M_i+\frac{1}{2}}\otimes\bigl|\bar
c\downarrow\bigr\rangle, \end{equation}  where
$C^{JM}_{j_1M_1;j_2M_2}$ are the corresponding Clebsch--Gordan
coefficients.

Later in the w.f. we neglect  possible (very small) mixing between
$D(1^-_{1/2})$,$D(1^-_{3/2})$ states and also between $D_s(2^+_{3/2})$,
$D_s(2^+_{5/2})$ states. However, physical $D_s(1^+)$ states can
be mixed via open channels and tensor interaction,while
the $0^+$ and $2^+$ levels are obtained
solely from $j=\frac12$ and $j=\frac32$, respectively.

The eigenstates, defining the higher  $1^+_H$ and lower $1^+_L$
levels, can be parametrized by introducing the  mixing angle
$\phi$: \begin{equation}|1^+_H\ran = \cos \phi |j=\frac12\ran +
\sin \phi|j=\frac32\ran\label{5},\end{equation}and
 \begin{equation}|1^+_L\ran = -\sin\phi |j=\frac12\ran + \cos\phi
|j=\frac32\ran,\label{6}\end{equation}where the mixing angle  is
 defined by the unitary mixing matrix $\hat O_{mix}$.
In the heavy-quark limit the states with $j=\frac32 $ and
$j=\frac12$ are not mixed, but for finite  $m_Q$  they can be
mixed and  definition of the   mixing matrix in this basis is
rather complicate procedure \cite{10I}, which is also
model-dependent. Therefore it is more convenient to connect the
angle $\phi$ in  (\ref{5}),(\ref{6}) with the  known factors in
the $\veL\veS$ basis, where $\hat O_{mix}$ is well defined in
closed-channel approximation and factually depends only on the
ratio $a/t$, where $a$ is the spin-orbit  and $t$ is the tensor
splitting. For our analysis we do not need to know details of
spin-orbit interaction (see Appendix C).

Then the splittings of the $2^+$ and $0^+$ levels are
\begin{equation} M(2^+) - M_{cog}=a-0.1 t, \label{7}\end{equation} \[M(0^+) -
M_{cog}=-2a- t, \] while $1^+_L$ and $1^+_H$  in  (\ref{5}) and
(\ref{6}) can be expressed through the mixing angle $\theta$ in
the expansion of these states in the $\veL\veS$ basis, where they
represent the decomposition of the $~^3P_1$ and $~^1P_1$ states:

\[| 1^{+}_H=\cos \theta  | ~^3P_1> - \sin\theta |~^1P_1>,\]
\begin{equation}|1^+_{L}= \sin\theta  | ~^3P_1> +\cos\theta |~^1P_1>
.\label{8}\end{equation}The states $^3P_1$ and $^1P_1$ in
(\ref{8}) can be expressed through the basis with the eigenstates
$|j=\frac32\ran$ and $j=\frac12\ran$ \cite{29I}:
\begin{equation}|~^3P_1\ran =\frac{1}{\sqrt{3}}|j=\frac32\ran +
\sqrt{\frac23} |j=\frac12\ran,\label{9}\end{equation} \[
|~^1P_1\ran =\sqrt{\frac23}|j=\frac32\ran -\frac{1}{\sqrt{3}}
|j=\frac12\ran.\]
 Then in closed-channel approximation the following
 relation can be established between angles $\phi$ and $\theta$:
 \begin{equation}\phi=-\theta + 35.264^{\circ}
 .\label{10}\end{equation} From (\ref{10}) it follows that
 \begin{enumerate}
    \item  If $1^+_H$ is pure $|j=\frac12>$ state ($\phi=0^{\circ})$, then this state is the admixture of the $~^3P_1$
    and $~^1P_1$ states with $\theta=35.264^{\circ}$.
    \item If  $1^+_H$ is pure $|j=\frac32>$ state ($\phi=90^{\circ})$, then in
the $\veL\veS$ basis this state is
    admixture of the $~^3P_1$ and $~^1P_1$ states with
    $\theta=-54.736^{\circ}$.
    \item The special case with   $\phi=-9.74^{\circ}$
     corresponds to "equal" mixing between the
     $~^3P_1$ and $~^1P_1$ states with  the angle
     $\theta=-45^{\circ}$.
 \end{enumerate}

 Therefore the solutions with small $|\phi|\lesssim 6^{\circ}$ correspond to
 the mixing angle $\theta$ from the range:
 $29^{\circ}\lesssim \theta\lesssim 41^{\circ}$ for which the ratio
 $a/t$ appears to be  slightly smaller (larger) unity for
 small negative (positive) $\phi$. Here we consider small negative
 $\phi$.

 The dependence of the mixing angle  on  the
 ratio $a/t$ is illustrated by Table \ref{alla_table_1} taking three   different
 ratios $a/t$.

\begin{table}
\caption{The mixing  coefficients  of the $1^+$ states for
different ratios $a/t$.} \label{alla_table_1}
\begin{center}
\begin{tabular}{|c|c|c|c|}\hline&&&\\
& $\frac{a}{t}=1.0$&  $\frac{a}{t}=0.950$& $\frac{a}{t}=0.917$\\
&$\theta=35.264^{\circ}$&$\theta=43.987^{\circ}$&$\theta=50.014^{\circ}$\\
&$\phi=0$&$\phi=-8.727^{\circ}$&$\phi=-14.75^O$\\\hline&&&\\
$1^+_H$& $+1.0|\frac12\ran $& $-0.152|\frac32\ran +
0.988|\frac12\ran$& -0.255$|\frac32\ran+0.967|\frac12\ran$\\&&&\\
\hline &&&\\ $1^+_L$& $1.0|\frac32\ran $& $0.988|\frac32\ran +
0.152|\frac12\ran$&0.255$|\frac32\ran + 0.967|\frac32\ran$\\&&&\\
\hline
\end{tabular}
\end{center}
\end{table}

Thus $1^+_H$ state as the pure $|j=\frac12>$ corresponds to $a=t$,
while for slightly smaller ratio, $a/t=0.95$, the admixture of the
$|j=\frac32>$ state is  $\sim 15\%$ and for $a/t=0.917$ the
admixture is already $26\%$. Notice that the physical condition
$a\lesssim t$ contradicts the heavy-quark limit when $t\to 0$
while $a\neq 0$ and can have large magnitude.

The structure of the mixing is important because it defines the
order of levels and the value of mass shift for the $1^{+'}$
state, as well as the mass shift and the width of another $1^+$
level. It is important that if the coupling to nearby continuum
channel is taken into account, then as follows from experiment,
the position of the $2^+$ and $1^+$ levels does not change (within
1-3 MeV) and just their mass difference $\Delta=M(2^+) - M(1^+)$
can be used to define tensor splitting: it can be derived that
$\Delta=(1.25\pm 0.15)t$ for any $a/t$.

\section{The masses of heavy-light mesons}

In closed-channel approximation the masses of HL mesons, or
initial positions of the levels (without channel coupling), can be
calculated in different schemes, e.g. in the $\veL\veS$ coupling
\cite{18I}, or as in the Dirac type coupling \cite{10I}. In Tables
\ref{alla_table_2}, \ref{alla_table_3} we give these unperturbed
masses for the $B$ and $B_s$ mesons which are calculated with the
use of the relativistic string Hamiltonian \cite{6I, 24I}. In this
approach the $P$-wave masses of HL mesons appear to be smaller
that in other potential models because they contain negative
string corections (see Appendix B).


\begin{table}
\caption{The $B$ meson masses (in MeV) (without decay channel
coupling)} \label{alla_table_2}
\begin{center}
\begin{tabular}{|c|c|c|c|c|c|c|}\hline&&&&&&\\
& $0^-$&  $1^-$& $0^+$&  $1^+_L$&$1^+_H$&$2^+$\\&&&&&&\\ \hline&&&&&&\\
From [20] and &5279&5325&5695&5726&5732&5742\\
this paper&&&&&&\\ \hline&&&&&&\\
experiment& 5279.0&5325.0&&5721$\pm5$ [28]&&5746$\pm4$[28]\\
&$\pm 0.5$[26]& $\pm0.6$[15]&& $5725.3^{+2.4}_{-3.2}$ [29]&&5739.9$^{+2.2}_{-2.4}$[29]\\
&&&&&&\\ \hline
\end{tabular}
\end{center}
\end{table}

Calculated masses of the states with $j=3/2$, $1^{+}_L$ and $2^+$,
for the ratio $a/t=0.95$ (see Table \ref{alla_table_2}) appear to
be in good agreement with recent DO Collab.measurements of the $B$
meson masses \cite{28I}. Such agreement can also be reached  for
other values  $a/t$ close to unity: \begin{equation}R =\frac{a}{t}
=1.0\pm 0.05,\label{11}\end{equation}and we take the same ratio
for the $B$, $B_s$ mesons, and also for the $D$,$D_s$ mesons: for
such choice the contribution from the $j=\frac12$ state dominates
in the $1^{+}_H$ meson. In particular for $a/t=0.95$:

\[ \left| 1^+_{H}\ran =0.9884\right| \frac12\ran -
\left.0.1517\right|\frac32\ran,\] and
\begin{equation}\left| 1^+_{L}\ran =0.1517\right|
\frac12\ran+\left.0.9884\right|\frac32\ran.\label{12}\end{equation}

The masses given in Tables \ref{alla_table_2}, \ref{alla_table_3}
are obtained taking the tensor splitting $t\cong 12.2$ MeV and
$t\cong 10$ MeV for the $B$ and $B_s$ mesons, respectively. The
tensor splittings have been determined to fit the mass difference
$M(2^+)-M(1^+)$, which has the important property --- it does not
change (within 2 MeV) if DCC is taken into account.


\begin{table}
\caption{The $B_s$ meson masses (in MeV) (without decay channel
coupling)} \label{alla_table_3}
\begin{center}
\begin{tabular}{|c|c|c|c|c|c|c|}\hline&&&&&&\\
 $J^P$& $0^-$&  $1^-$& $0^+$&  $1^+_L$&$1^+_H$&$2^+$\\&&&&&&\\ \hline&&&&&&\\
This paper&5362&5407&5805&5830&5835&5843\\
and from [20]&&&&&&\\ \hline&&&&&&\\
experiment& 5367.7&5411.7&&5829.4&&5839.1\\
&$\pm 1.8$[26]& $\pm3.2$[30]&& $\pm0.8$[28]&&$\pm3.0$[28]\\
&&&&&&\\ \hline
\end{tabular}
\end{center}
\end{table}

In Table \ref{alla_table_4} the masses  of the $B,B_s$ mesons
(from Tables \ref{alla_table_2},\ref{alla_table_3}) are compared
to those obtained in other models; there  the conventional
notations $1^+$ and $1^{+'}$ for the $j=\frac32$ and $j=\frac12$
states are used.


\begin{table}
\caption{Theoretical predictions for the $B(1P)$ and $B_s(1P)$
masses (in MeV)(without decay channel coupling)}
\label{alla_table_4}
\begin{center}
\begin{tabular}{|c|c|c|c|c|c|c|}\hline&&&&&&\\
Ref.& [3]&[4]&[10]&[14]& This paper&exp.\footnotemark[1]\footnotemark[2]\\&&&&&&\\
\hline&&&&&&\\
$M_B(0^+)$&5760&5738&5706&5700&5695 (10)& abs\\
&&&&&&\\
$M_B(1^{+'})$&5780&5757&5742&5750&5732& abs\\
&&&&&&\\
$M_B(1^+)$&5780&5719&5700&5774&5726&5721(5)\footnotemark[1]\\&&&&&&5725(3)\footnotemark[2]\\
$M_B(2^+)$&5800&5733&5714&5790&5742&5746(4)\footnotemark[1]\\&&&&&&5740(2)\footnotemark[2]\\
\hline&&&&&&\\
$M_{B_s}(0^+)$&5830&5841&5804&5710&5805(10)& abs\\
&&&&&&\\
$M_{B_s}(1^{+'})$&5860&5859&5842&5770&5835& abs\\
&&&&&&\\ $M_{B_s}(1^+)$ &5860&5831&5805&5870&5830&5829(1)\footnotemark[2]\\&&&&&&\\
$M_{B_s}(2^+)$ &5888&5844&5820&5893&5843&5840(1)\footnotemark[2]\\&&&&&&\\
 \hline
\end{tabular}
\end{center}
\footnotetext[1]{Experimental data of the D0 Collaboration [27]}
\footnotetext[2]{Experimental data  from [28]}
\end{table}



Comparison of  the  masses,given in Table \ref{alla_table_4},
shows that in different papers $M(B), M(B_s)$  differ not much,
within $\pm 50$ MeV, however, the order of levels inside the 1P
multiplet appears to be different.In particular, in \cite{4I},
\cite{10I} the $2^+$ level has smaller mass than the $1^{+'}$
while  in our calculations the $2^+$ state has always maximal
mass.It means that FS parameters $a,t,$ and their ratio, as well
as  the mixing angle $\phi$ between the $1^+$ and $1^{+'}$ states,
can differ essentially  in given papers. Meanwhile existing
experimental limit on the width of $D_{s1}(2536)$ puts strict
restrictions on admittable value of the mixing angle $\phi$ (see
next Section).

Finally in Table \ref{alla_table_5} we give also unperturbed
masses of the $D(1P)$ and $D_s(1P)$ mesons, taking the splitting
$t=29$ MeV from the mass difference, $\Delta=M(2^+) -M(1^+)=1.31
t= 38$ MeV and $a/t=0.95$, both for the $D$ and $D_s$
mesons.Notice that the position of the $P$-wave mesons does not
practically change if $a/t=1.0$ (or $\phi=0$).

\begin{table}
\caption{The  masses of the $D(1P)$ and $D_s(1P)$ mesons in MeV)
(without decay channel coupling)} \label{alla_table_5}
\begin{center}
\begin{tabular}{|l|l|l|l|l|}\hline &&&&\\
&$0^+$&$1^+_L$&$1^+_H$&$2^+$\\
 \hline &&&&\\
$D$&2352&2423&2436&2461\\&&&&\\\hline experiment&
2350&2422&2427&2459\\
$[26]$&$\pm 50$&$\pm 1.3$&$\pm 51$&$\pm 4$\\
 &&&&\\\hline
$D_s$&2467&2537&2550&2575\\
&&&&\\\hline experiment&
2317.3&2535.4&2459&2573.5\\
$[26]$&$\pm 0.6 $&$\pm 0.8$&$\pm 1$&$\pm 1.7$\\
 &&&&\\\hline
\end{tabular}
\end{center}
\end{table}

Given in Table \ref{alla_table_5} masses  show that in
closed-channel approximation we have reached good agreement with
experiment for all $D(1P)$ mesons and also for narrow mesons
$D_s(2535), D_s(2573)$.

We do  not  need here to know the details of spin-orbit interaction
which at present is not fully understood, probably, because of
important role of one-loop (or even higher) corrections \cite{31I} and
possible suppression of NP spin-orbit potential observed on the
lattice \cite{32}. Here we would like only to notice that in heavy
quarkonia the ratios $a/t$ are also close or equal  unity:

\[
a/t=1.04\pm 0.08 ~~ {\rm~ for}~ \chi_b(1P);~~ a/t=(1.02\pm 0.14)
 {\rm~ for}~ \chi_b(2P),\]
\begin{equation}a/t=(0.86\pm 0.02)  {\rm~ for}~ \chi_c(1P).\label{13}\end{equation}

\section{Chiral Transitions}

 To obtain the mass shift due to DCC effect we use here
the chiral Lagrangian (1), which includes both effects of
confinement (embodied in the string tension) and Chiral Symmetry
Breaking (CSB)  (in Euclidean notations): \begin{equation}L_{FCM}
= i\int d^4 x \psi^+ (\hat\partial +m+ \hat M)
\psi\label{14}\end{equation}with the mass operator $\hat M$ given
as a product of the scalar function $W(r)$ and the  SU(3) flavor
octet,

\begin{equation}\hat M = W(r) \exp (i\gamma_5
\frac{\varphi_a\lambda_a}{f_\pi}),\label{15}\end{equation}where
\begin{equation}\varphi_a\lambda_a =\sqrt2
\left(\begin{array}{lll}
\frac{\pi^0}{\sqrt{2}}+\frac{\eta^0}{\sqrt{6}},& \pi^+,& K^+\\
\pi^-,& \frac{\eta^0}{\sqrt{6}}-\frac{\pi^0}{\sqrt{2}},& K^0\\
K^-, &\bar K^0,&
-\frac{2\eta^0}{\sqrt{6}}\end{array}\right).\label{16}\end{equation}Taking
the meson emission to the lowest order, one obtains the quark-pion
Lagrangian in the form \begin{equation}\Delta L_{FCM} =- \int
\psi_i^+ (x) \sigma |\vex|\gamma_5 \frac{\varphi_a\lambda_a}{f\pi}
\psi_k(x)d^4x.\label{17}\end{equation}

Writing the equation (\ref{17}) as $\Delta L_{FCM}=- \int V_{if}
dt$, one obtains the operator matrix element for the transition
from the light quark state $i$ (i.e. the initial  state $i$ of a
HL meson) to the continuum state $f$ with the emission of a NG
meson $(\varphi_a\lambda_a)$. Thus we are now able to write the
coupled channel equations, connecting any state of a HL meson  to
a decay channel  which contains another HL meson plus a NG meson.

In  the case, when interaction in each channel and also  in the
transition operator is time-independent, one can write following
system of equations (see \cite{33} for a review)
\begin{equation}[(H_i -E) \delta_{il} +V_{il}]
G_{lf}=1.\label{18}\end{equation}Such two-channel system of the
equations can be reduced to one equation with additional DCC
potential, or the Feshbach potential
\cite{34},\begin{equation}(H_1-E) G_{11} -V_{12} \frac{1}{H_2-E}
V_{21} G_{11}=1.\label{19}\end{equation}Considering a complete set
of  the states $|f\ran$ in the  decay channel 2 and the set of
unperturbed states $|i\ran$ in channel 1, one arrives at the
nonlinear equation for the shifted mass $E$,
\begin{equation}E=E_1^{(i)} - \sum_f \lan i|V_{12} | f\ran
\frac{1}{E_2^{(f)}-E} \lan f
|V_{21}|i\ran.\label{20}\end{equation} Here  the unperturbed
values of $E^{(i)}_1$ are assumed to be known (see Tables
\ref{alla_table_2}, \ref{alla_table_3}, \ref{alla_table_5}), while
the interaction $U_{if}$ is defined in (\ref{17}). A solution of
the nonlinear equation (\ref{20}) yields (in general a complex
number $E=\bar E-\frac{i\Gamma}{2})$ one or more roots on all
Riemann sheets of the complex mass plane.


\section{Calculation of the DCC shifts}

To calculate explicitly the  mass shifts, we will use  the Eq.
(\ref{20}) in the following form:
\begin{equation}
m[i]=m^{(0)}[i]-\sum\limits_f \dfrac{|<i|\Hat V|f>|^2}{E_f-m[i]},
\label{21}
\end{equation}
where $m^{(0)}[i]$ is the initial mass, $m[i]$ -- is the final
one, $E_f= \omega_D+\omega_K$ is the energy of the final state,
and the operator $\hat V$ provides the transitions between the
channels (see the comment after Eq. (\ref{17})).

In our approximation we do not take into account the final state
interaction in the $DK$ system and neglect the $D$-meson motion,
so the w.f. of the $i,f$-states are:
\begin{equation}
|f>=\Psi_K(\pp)\otimes\Psi_{D}(M_f),\quad |i>=\Psi_{D_s}(M_i),
\end{equation}
where
\begin{equation}
\Psi_K(\pp)=\frac{e^{i\pp\rr}}{\sqrt{2\omega_K(\pp)}}
\end{equation}
is the plane wave describing the $K$-meson and
$\Psi_{D}(M_f)$, $\Psi_{D_s}(M_i)$ are the HL meson w.f. at rest
with the spin projections $M_f$, $M_i$, respectively.

We introduce the following notations:
\begin{equation}
\omega_K(\pp)=\sqrt{\pp^2+m_K^2},\quad
\omega_D(\pp)=\sqrt{\pp^2+m_D^2},
\end{equation}
so that in the final state the total energy is
$E_f=\omega_D+\omega_K$, while
\begin{equation}
T_f=E_f-m_D-m_K
\end{equation}
is the  kinetic energy. Also it is convenient to define other
masses with respect to nearby threshold: $m_{thr}= m_K+m_D$,
\begin{equation}
E_0=m^{(0)}[D_s]-m_D-m_K,\quad \delta m=m[D_s]-m^{(0)}[D_s],
\end{equation}
\begin{equation}
\Delta = E_0+\delta m=m[D_s]-m_D-m_K,
\end{equation}
where $\Delta$ determines the deviation of the $D_s$ meson mass
from the threshold. In what follows we consider unperturbed masses
$m_0(J^P)$ of the ($Q\bar q$) levels as given (our results do not change
if we slightly vary their position, in this way the analysis is actually
model-independent).

Using these notations, the Eq.(\ref{20}) can be rewritten
as
\begin{equation}
E_0-\Delta=\xxi(\Delta), \label{misha_eq_10}
\end{equation}
where
\begin{equation}
\xxi(\Delta)\overset{\text{def}}{=}
\int\frac{d^3\pp}{(2\pi)^3}\sum\limits_{M_f}
\frac{\left|\left\langle M_i\left|\hat
V\right|\pp,M_f\right\rangle\right|^2}{T_f(\pp)-\Delta}
\end{equation} and
\begin{equation}
\left\langle M_i\left|\hat V\right|\pp,M_f\right\rangle=-\int
\Psi^\dag_{D_s}(M_i)\,\sigma
|\rr|\gamma_5\frac{\sqrt{2}}{f_K}\,\Psi_{D}(M_f)\,
\frac{e^{i\pp\rr}}{\sqrt{2\omega_K(\pp)}}\, d^3\rr,
\label{misha_eq_30}
\end{equation}

The function $\xxi(\Delta)$ for negative $\Delta$ diminishes
monotonously so there exists a final (critical) point,
\begin{equation}
E_0^{\text{crit}}=\xxi(-0).
\end{equation}
Thus, while solving the Eq.(\ref{misha_eq_10}), one has two
possible situations: $E_0<E_0^{\text{crit}}$ and
$E_0>E_0^{\text{crit}}$.

\begin{figure}
 \caption{Eq.(\ref{misha_eq_10}) for the case
$E_0<E_0^{\text{crit}}$} \label{misha_fig_1}
\includegraphics[width=100mm,keepaspectratio=true]{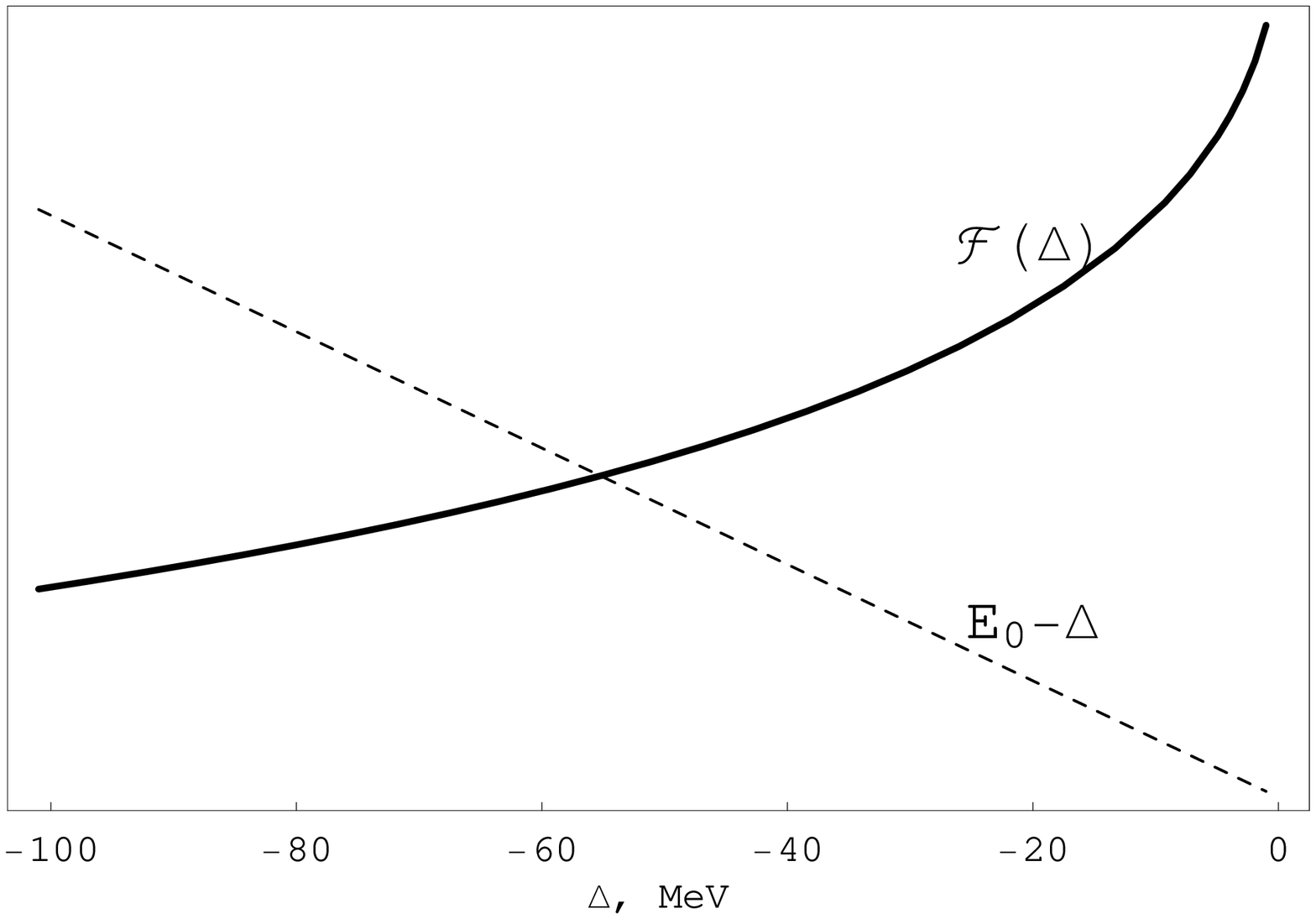}
\end{figure}

\begin{figure}
 \caption{Eq.(\ref{misha_eq_10}) for the case
$E_0>E_0^{\text{crit}}$} \label{misha_fig_2}
\includegraphics[width=100mm,keepaspectratio=true]{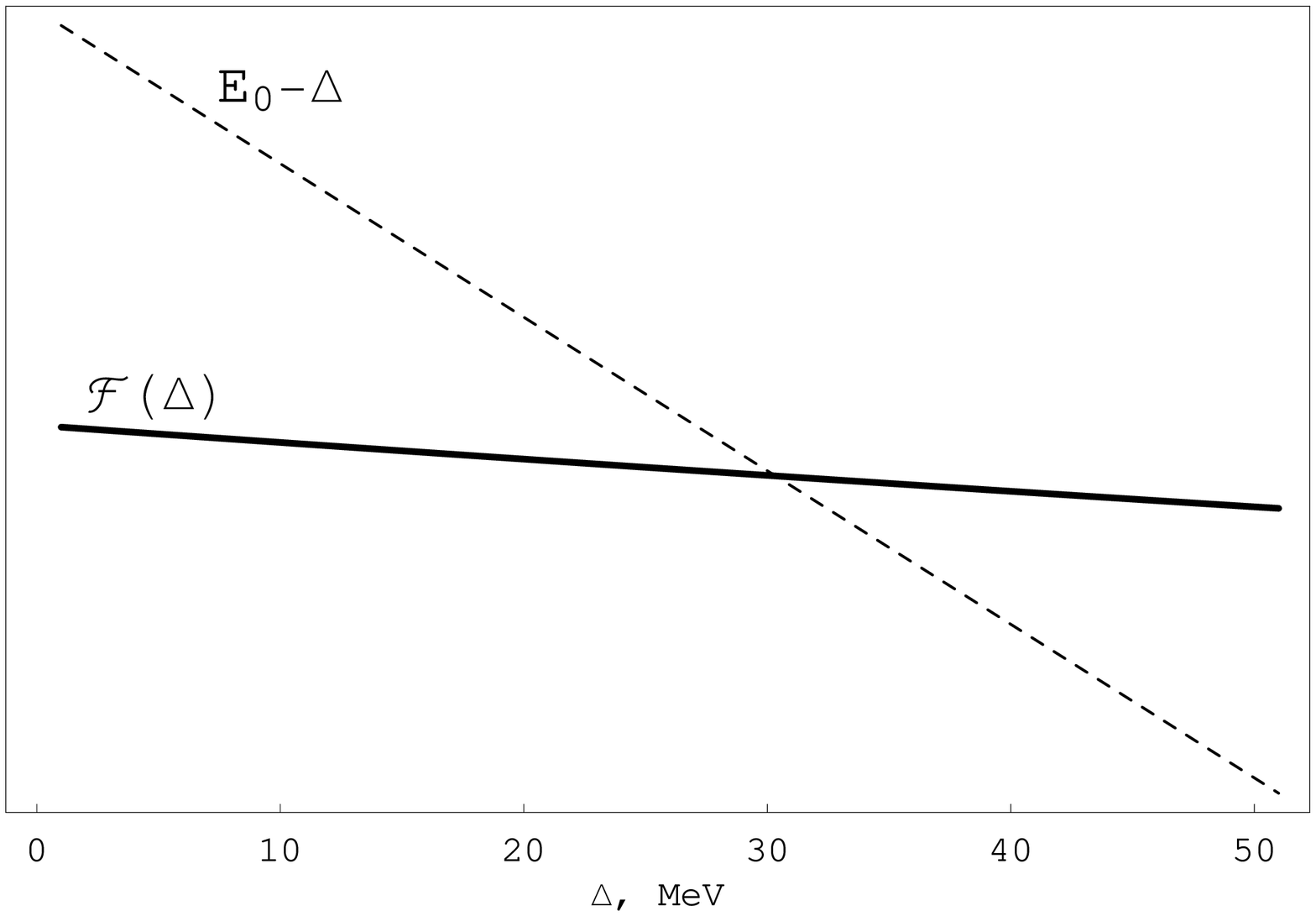}
\end{figure}

In the first case Eq.(\ref{misha_eq_10}) has a negative real root
$\Delta<0$ (see Fig. \ref{misha_fig_1}) and the resulting mass of
the $D_s$ meson appears to be under the threshold. In the second
case Eq.(\ref{misha_eq_10}) has a complex root
$\Delta=\Delta'+i\Delta''$ with positive real part $\Delta'>0$
(see Fig. \ref{misha_fig_2}) and negative imaginary part
$\Delta''<0$. To find latter solutions one should  make analytic
continuation of the solution(s) from the upper halfplane of
$\Delta$ under the cut, which starts at the threshold, to the
lower halfplane (second sheet). This solution can be also obtained
by deforming the integration contour in $T_f(p)$. In actual
calculations we take infinitesimal imaginary part $\Delta''$,
proving that $\Delta$ does not change much for finite $\Delta''$
(the similar procedure has been used in \cite{18I}). Finally, the
resulting mass of the $D_s$ meson proves to be in the complex
plane at the position $\Delta'-i|\Delta''|$, i.e. the meson has
the finite width $\Gamma=2\Delta''$.

For further calculations  we should insert the explicit meson w.f.
to the matrix element (\ref{misha_eq_30}). As discussed above,
in a HL meson we consider a light quark $q$ moving in
the static field of a heavy antiquark $\bar Q$, and therefore
its w.f. can be taken  as the Dirac bispinor:
\begin{equation} \psi_q^{jlM}=\begin{pmatrix}
g(r)\Omega_{jlM}\\
(-1)^{\frac{1+l-l'}{2}}f(r)\Omega_{jl'M}\end{pmatrix},\quad
\int\limits_0^\infty \left(f^2+g^2\right)r^2dr=1,\end{equation}
where the functions $g(r)$ and $f(r)$ are the  solutions of the
Dirac equation: \begin{equation}
\begin{gathered} g'+\frac{1+\varkappa}{r}g-\left(E_q+m_q+U-V_C\right)f=0,\\
f'+\frac{1-\varkappa}{r}f+\left(E_q-m_q-U-V_C\right)g=0.
\end{gathered} \end{equation} Here the interaction between the quark
and the antiquark is described by a sum of  linear scalar
potential and the  vector Coulomb potential with
$\alpha_s=\text{const}$:
\begin{equation}
U=\sigma r,\quad V_C=-\dfrac{\beta}{r},\quad
\beta=\dfrac{4}{3}\alpha_s.
\end{equation}

Introducing new dimensionless variables
\begin{equation}
x=r\sqrt{\sigma},\quad \varepsilon_q=E_q/\sqrt{\sigma},\quad
\mu_q=m_q/\sqrt{\sigma}, \end{equation} and new dimensionless
functions
\begin{equation} g=\sigma^{3/4}\frac{G(x)}{x},\quad
f=\sigma^{3/4}\frac{F(x)}{x},\quad \int\limits_0^\infty
\left(F^2+G^2\right)dx=1, \end{equation} we come to the following
system of equations:
\begin{equation}
\begin{gathered}
G'+\frac{\varkappa}{x}G-\left(\varepsilon_q+\mu_q+x+\frac{\beta}{x}\right)F=0,
\\
F'-\frac{\varkappa}{x}F+\left(\varepsilon_q-\mu_q-x+\frac{\beta}{x}\right)G=0.
\end{gathered}
\end{equation}This system has been solved numerically.

\begin{figure}
\caption{$G_{1,2,3}(x)$ functions} \label{misha_fig_3}
\includegraphics[width=100mm,keepaspectratio=true]{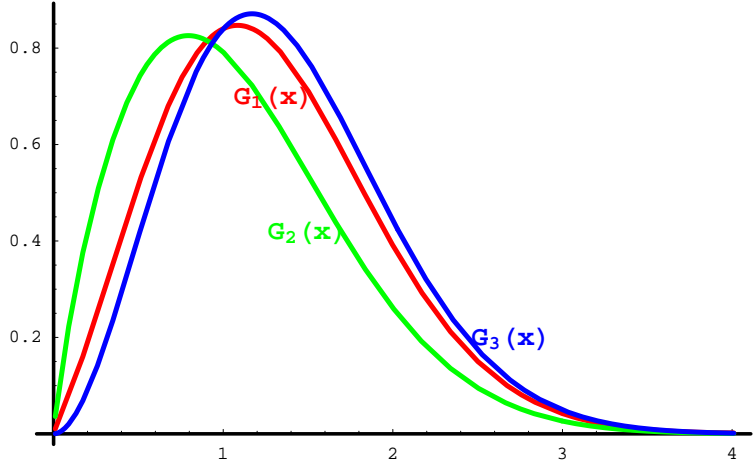}
\end{figure}

\begin{figure}
\caption{$F_{1,2,3}(x)$ functions} \label{misha_fig_4}
\includegraphics[width=100mm,keepaspectratio=true]{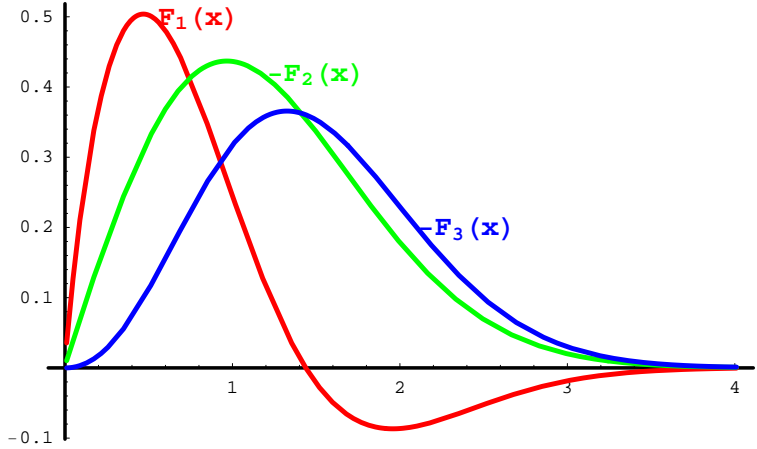}
\end{figure}

Using the parameters from the papers \cite{18I}, \cite{dirac}:
\begin{equation} \begin{gathered} \sigma=0.18~\text{GeV}^2,\quad \alpha_s=0.39,
\\ m_s=210~\text{MeV},\quad m_q=4~\text{MeV}, \end{gathered}
\end{equation}
we obtain the following Dirac eigenvalues $\varepsilon$:
\begin{equation}
\begin{tabular}{r|l|l} $\varkappa$ &
$\bar Qq$, $\mu_q=0.01$~ & $\bar Qs$, $\mu_s=0.5~$ \\
\hline\hline -1 & 1.0026 & 1.28944 \\ \hline +1 & 1.7829 & 2.08607 \\
\hline -2 & 1.7545 & 2.08475
\end{tabular}
\label{misha_table_2}
\end{equation}
and corresponding eigenfunctions $G$, $F$ are given in Figs.
\ref{misha_fig_3}, \ref{misha_fig_4}).

Later we use the simplified notations for the
quark bispinors:
\begin{equation}
\psi_1(M_1)\overset{\text{def}}{=}\psi_s^{\frac{1}{2},1,M_1},\quad
\psi_2(M_2)\overset{\text{def}}{=}\psi_q^{\frac{1}{2},0,M_2},\quad
\psi_3(M_3)\overset{\text{def}}{=}\psi_s^{\frac{3}{2},1,M_3}.
\end{equation}

\begin{figure}
\caption{$\Phi_{0,2}(q)$ functions} \label{misha_fig_5}
\includegraphics[width=100mm,keepaspectratio=true]{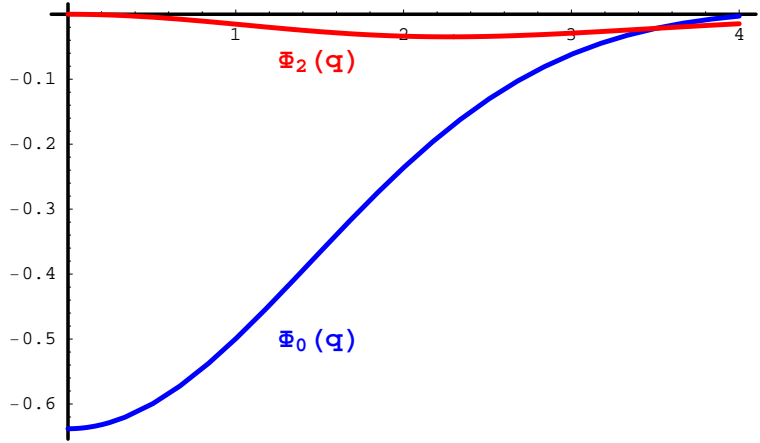}
\end{figure}

Now, using  explicit expressions for the spherical spinors,
\begin{equation}\Omega_{l+1/2,l,M}=\begin{bmatrix}
\sqrt{\frac{j+M}{2j}}\,Y_{l,M-1/2} \vphantom{\bigg|} \\
\sqrt{\frac{j-M}{2j}}\,Y_{l,M+1/2} \vphantom{\bigg|}
\end{bmatrix},\quad \Omega_{l-1/2,l,M}=\begin{bmatrix}
-\sqrt{\frac{j-M+1}{2j+2}}\,Y_{l,M-1/2} \vphantom{\bigg|} \\
\sqrt{\frac{j+M+1}{2j+2}}\,Y_{l,M+1/2} \vphantom{\bigg|}
\end{bmatrix},\end{equation}
and the expansion :
\begin{equation} e^{i\pp\rr}=4\pi\sum\limits_{l,M} i^l j_l(pr)
Y^*_{l,M}\left(\frac{\pp}{p}\right)
Y_{l,M}\left(\frac{\rr}{r}\right)\, , \end{equation} after
cumbersome transformations (which are omitted in the text) we
obtain the transition matrix elements:
\begin{multline} \Bigl\|\mathcal{V}_{12}\Bigr\|_{M_1,M_2}=-\int
\psi^\dag_1(M_1)\,\sigma
|\rr|\gamma_5\frac{\sqrt{2}}{f_K}\,\psi_2(M_2)\,
\frac{e^{i\pp\rr}}{\sqrt{2\omega_K(\pp)}}\, d^3\rr=\\
=\frac{\sqrt{\sigma}}{f_K\sqrt{\omega_K(p)}}\Phi_0\left(\frac{p}{\sqrt{\sigma}}\right)
\sqrt{4\pi}Y^*_{0,M_1-M_2}\left(\frac{\pp}{p}\right),
\end{multline}
\begin{multline} \Bigl\|\mathcal{V}_{32}\Bigr\|_{M_3,M_2}=-\int
\psi^\dag_3(M_3)\,\sigma
|\rr|\gamma_5\frac{\sqrt{2}}{f_K}\,\psi_2(M_2)\,
\frac{e^{i\pp\rr}}{\sqrt{2\omega_K(\pp)}}\, d^3\rr=\\
=-\frac{\sqrt{\sigma}}{f_K\sqrt{\omega_K(p)}}\Phi_2\left(\frac{p}{\sqrt{\sigma}}\right)
\sqrt{\frac{4\pi}{5}}Y^*_{2,M_3-M_2}\left(\frac{\pp}{p}\right)
\cdot \begin{bmatrix} -1 & +2 \\ -\sqrt{2} & +\sqrt{3} \\
-\sqrt{3} & +\sqrt{2} \\ -2 & +1
\end{bmatrix}\, .
\end{multline}
where
\begin{equation} \begin{gathered} \Phi_0(q)=\int\limits_0^\infty j_0(qx)xdx\Bigl[
G_1(x)F_2(x)-F_1(x)G_2(x) \Bigr], \\
\Phi_2(q)=\int\limits_0^\infty j_2(qx)xdx\Bigl[
G_3(x)F_2(x)-F_3(x)G_2(x) \Bigr]. \end{gathered} \end{equation}
Notice that because of different signs of the $F_1(x)$ and
$F_{2,3}(x)$ functions (while the $G_{1,2,3}$ functions are all
positive) on almost all real axis, the integral $\Phi_2$ appears
to be strongly suppressed in comparison with the integral
$\Phi_0$. This fact is confirmed by numerical simulations (see
Fig. \ref{misha_fig_5}).

Finally, introducing universal functions
\begin{equation}
\begin{gathered}
\tilde\xxi_{0,2}(\Delta)=\frac{\sigma}{2\pi^2f_K^2}\int\limits_0^\infty
\frac{p(T_f)\omega_D(T_f)dT_f}{T_f+m_D+m_K}\cdot
\frac{\Phi_{0,2}^2\left(\dfrac{p(T_f)}{\sqrt{\sigma}}\right)}{T_f-\Delta}\,
,
\\ \tilde\Gamma_{0,2}(T_f)=\frac{\sigma}{\pi f_K^2}\cdot
\frac{p(T_f)\omega_D(T_f)}{T_f+m_D+m_K} \cdot
\Phi_{0,2}^2\left(\frac{p(T_f)}{\sqrt{\sigma}}\right)\, ,
\end{gathered}
\end{equation}
we come to the following equations to determine meson masses and
widths:
\begin{equation}
\begin{array}{rc||cl} D_s(0^+) \vphantom{\bigg|} &
\hphantom{a} & \hphantom{a} &
E_0[0^+]-\Delta=\tilde\xxi_0(\Delta),
\\ D_s(1^+_L) \vphantom{\bigg|} & & & E_0[1^+_L]-\Delta=
\cos^2\phi\cdot\tilde\xxi_0(\Delta)+\sin^2\phi\cdot\tilde\xxi_2(\Delta),
\\ D_s(1^+_H)
\vphantom{\bigg|} & & & E_0[1^-_H]-\Delta'=
\sin^2\phi\cdot\tilde\xxi_0(\Delta')+\cos^2\phi\cdot\tilde\xxi_2(\Delta'),
\\ \vphantom{\bigg|} & & &
\Gamma[1^+_H]=\sin^2\phi\cdot\tilde\Gamma_0(\Delta')
+\cos^2\phi\cdot\tilde\Gamma_2(\Delta'),
\\ D_s(2^+_{3/2})
\vphantom{\bigg|} & & & E_0[2^+_{3/2}]-\Delta'=
\dfrac{3}{5}\cdot\tilde\xxi_2(\Delta'),
\\ \vphantom{\bigg|} & & &
\Gamma[2^+_{3/2}]=\dfrac{3}{5}\cdot\tilde\Gamma_2(\Delta').
\end{array}
\label{misha_table_4}
\end{equation}

\section{Results and discussion}

In this chapter, using the expressions (\ref{misha_table_4}) to
define the $D_s$ and $B_s$ meson mass shifts, we present and
discuss our results. We will take into account the following pairs
of mesons in coupled channels ($i$ refers to first
(initial)channel, while $f$ refers to second (decay) one):

\begin{equation}
\begin{array}{c}
\begin{tabular}{c||c|c|c} $i$ & $D_s(0^+)$ &
$D_s(1^+)$ & $D_s(2^+)$
\\ \hline $f$ & $D(0^-)+K(0^-)$ & $D^*(1^-)+K(0^-)$ &
$D^*(1^-)+K(0^-)$
\end{tabular}\\[7mm]
\begin{tabular}{c||c|c|c} $i$ & $B_s(0^+)$ &
$B_s(1^+)$ & $B_s(2^+)$
\\ \hline $f$ & $B(0^-)+K(0^-)$ & $B^*(1^-)+K(0^-)$ &
$B^*(1^-)+K(0^-)$
\end{tabular}
\end{array}
\label{misha_table_1}
\end{equation}

In our calculations we use here the following meson
masses and thresholds (in MeV):
\begin{equation}
\begin{gathered}
m_{D^+}=1869, \quad m_{D^+}+m_{K^-}=2363, \\
m_{D^{*+}}=2010, \quad m_{D^{*+}}+m_{K^-}=2504,\\
m_{B^+}=5279, \quad m_{B^+}+m_{K^-}=5772, \\
m_{B^*}=5325, \quad m_{B^*}+m_{K^-}=5819. \\
\end{gathered}
\end{equation}

\begin{table}
\caption{$D_s(0^+)$-meson mass shift due to the $DK$ decay channel
and $B_s(0^+)$-meson mass shift due to the $BK$ decay channel (all
in MeV)} \label{misha_table_11}
\begin{center}
\begin{tabular}{|c|c|c|c|c|} \hline state  & $m^{(0)}$ &
$m^{\text{(theor)}}$ & $m^{\text{(exp)}}$ & $\delta m$  \\
\hline \hline $D_s(0^+)$  & 2467 & 2331 & 2317 & -136
\\ \hline $B_s(0^+)$ & 5805 & 5700 & {\footnotesize not seen} & -105 \\ \hline
\end{tabular}
\end{center}
\end{table}

\begin{table}
\caption{The $D_s(1^+)$, $D_s(2^+)$ meson mass shifts and widths
due to the $D^*K$ decay channel for the mixing angle $4^\circ$
(all in MeV)} \label{misha_table_12}
\begin{center}
\begin{tabular}{|c|c|c|c|c|c|c|} \hline state & $m^{(0)}$ &
$m^{\text{(theor)}}$ & $m^{\text{(exp)}}$ &
$\Gamma^{\text{(theor)}}_{(D^*K)}$ & $\Gamma^{\text{(exp)}}_{(D^*K)}$ & $\delta m$ \\
\hline \hline $D_s(1^+_H)$ & 2550 & 2440 & 2460 & $\times$ &
$\times$ & -110
\\ \hline $D_s(1^+_L)$ & 2537 & 2535 & 2535 & 1.1 & $<1.3$ & -2  \\
\hline $D_s(2^+_{3/2})$ & 2575 & 2573 & 2573 & 0.03 & {\footnotesize not seen} & -2\\
\hline
\end{tabular}
\end{center}
\end{table}

\begin{table}
\caption{The $B_s(1^+)$, $B_s(2^+)$ meson mass shifts and widths
due to the $B^*K$ decay channel for the mixing angle $4^\circ$
(all in MeV)} \label{misha_table_13}
\begin{center}
\begin{tabular}{|c|c|c|c|c|c|c|} \hline state & $m^{(0)}$ &
$m^{\text{(theor)}}$ & $m^{\text{(exp)}}$ &
$\Gamma^{\text{(theor)}}_{(B^*K)}$ & $\Gamma^{\text{(exp)}}_{(B^*K)}$ & $\delta m$ \\
\hline \hline $B_s(1^+_H)$ & 5835 & 5727 & {\footnotesize not
seen} & $\times$ & $\times$ & -108
\\  \hline $B_s(1^+_L)$ & 5830 & 5828 & 5829 & 0.8 & $<2.3$ & -2 \\
\hline $B_s(2^+_{3/2})$ & 5842 & 5840 & 5840 & $<10^{-3}$ & {\footnotesize not seen} & -2 \\
\hline
\end{tabular}
\end{center}
\end{table}

\begin{table}
\caption{The mixing coefficients  in Eq.(\ref{20}) for the DCC
shifts of  the $B_s, D_s$ mesons ($\phi=5.7^{\circ}$).}
\label{misha_table_14}
\begin{center}
\begin{tabular}{|c|c|c|c|}\hline &&&\\
&& $1^+_H$& $1^+_L$\\ &$a/t$&$\cos^2\phi$&$\sin^2\phi$\\ \hline &&&\\
$D_s(1P),B_s(1P)$&0.96& (0.995)$^2$& (0.100)$^2$\\ &&&\\\hline &&&\\
$D_s(1P),B_s(1P)$&1.0&1.0& 0\\ &&&\\\hline
\end{tabular}
\end{center}
\end{table}

The results of our calculation are presented in Tables
\ref{misha_table_11}--\ref{misha_table_13}. \textit{A priori} one
cannot say whether the $|j=\frac12\ran$ and $|j=\frac32\ran$
states are mixed or not.In \cite{35} in the case when there is no
mixing   at all, the width $\Gamma(D_{s1}(2536))= 0.3$ MeV is
obtained, while now the experimental limit is $\Gamma<2.3$ MeV
\cite{26I} and recently in \cite{36} the width $\Gamma=1.0\pm
0.17$ MeV has been measured. Therefore small mixing is not
excluded and here we take the mixing angle $\phi$ slightly
deviated from  $\phi=0^{\circ}$ (when there is no mixing at all).
Then one can define those angles $\phi$ which are compatible with
experimental data for the  masses and widths of both $1^+$ states.
The limiting angle $|\phi|= 5.7^{\circ}$ (given in Table
\ref{misha_table_14}) corresponds to the mixing between the
$^3P_1$ and $^1P_1$ with $\theta=41^{\circ}$ in $\veL\veS$ scheme.

\begin{figure}
\caption{Scheme of $D_s(1^+,2^+)$ shifts due to chiral coupling}
\label{misha_fig_6}
\includegraphics[width=100mm,keepaspectratio=true]{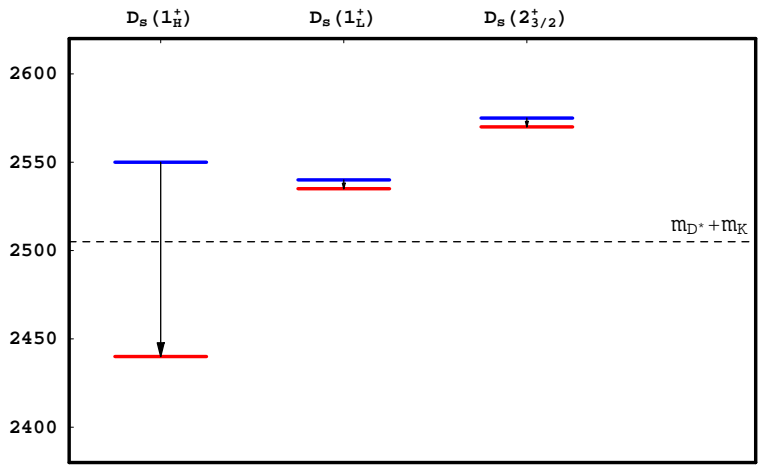}
\end{figure}

\begin{figure}
\caption{Scheme of $B_s(1^+,2^+)$ shifts due to chiral coupling}
\label{misha_fig_7}
\includegraphics[width=100mm,keepaspectratio=true]{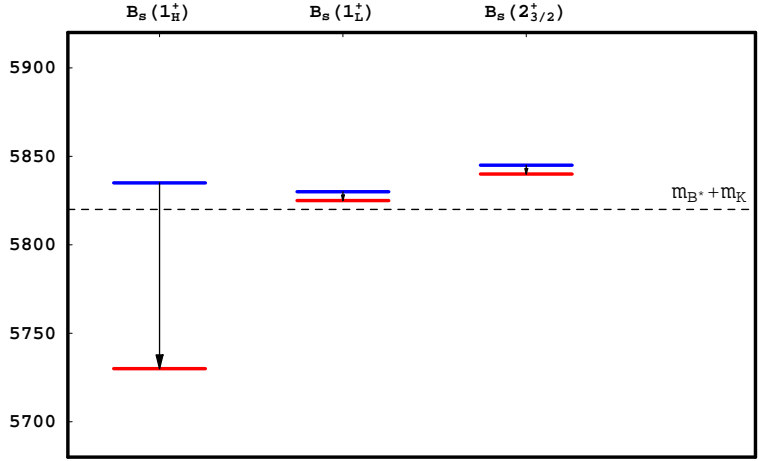}
\end{figure}

The value $\xi=(0.995)^2$ for the $1_{H}^+$ states provides large
mass shift ($\sim 100$ MeV) of this level and at the same time
does not produce the  mass shift of the $1^{+}_L$ level, which is
almost pure $j=\frac32$ state. For illustration we show the scheme
of the $1^+$, $2^+$ shifts on Figures \ref{misha_fig_6},
\ref{misha_fig_7}. We would like to stress here that the
dependence of the shift on the heavy non-strange meson mass (or,
equivalently, on the heavy quark mass) is rather weak, which
follows directly from the Eq. (\ref{misha_table_4}) via inverse
mass expansion.

If one uses  one-gluon  exchange potential for  tensor
interaction, then to obtain the splitting $t\sim 12$ MeV for the
$B$ and $B_s$ mesons and $t=29$ MeV for the $D$ and $D_s$ mesons
one needs to take $\alpha_s(\mu_{FS})\sim 0.39$ for all $P$-wave
$HL$ mesons. However, in OGE approximation the spin-orbit
splitting does not satisfy the condition $a=0.95t$  for this value
of $\alpha_s$.  This fact possibly indicates on important role  of
one-loop or even higher radiative corrections, observed in heavy
quarkonia \cite{3I}, and also on possible $\sim 20 \% $
 suppression of NP spin-orbit confining potential observed on the lattice
 \cite{32}.

\section{Conclusions}

We have studied the mass shifts of the $D_s(0^+, 1^{+'})$ and
$B_s(0^+, 1^{+'})$ mesons due to strong coupling to the decay
channels $DK, D^*K$ and $BK, B^*K$. To this end the chiral
quark-pion Lagrangian without fitting parameters has been used.

We have shown that the emission of a NG meson, accompanied with
the $\gamma_5$ factor, gives rise to maximal overlapping  between
the higher component with $j=\frac12$ of the $P$-wave meson
($D_s,B_s$) bispinor w.f. and the lower component (also with
$j=\frac12$) of the $S$-wave meson w.f. in considered $S$-wave
decay channel. Due to this effect, while taking the w.f. of the
$1P$ and $1S$ states with the use of the Dirac equation, large
mass shifts of the $0^+, 1^{+'}$ states are obtained.

The widths of $D_{s1}(2536)$ and $B_{s1}(5830)$ are also
calculated. To satisfy the experimental condition
$\Gamma(D_{s1}(2536))<2.3$ MeV the following limit on the mixing
angle $\phi$ (between the $|j=\frac32>$ and $|j=\frac12>$ states)
is obtained, $|\phi|\lesssim 6^{\circ}$. This restriction implies
that the mixing angle $\theta$ between the $^3P_1$  and $^1P_1$
states in $\veL\veS$ basis lies in the range: $29^{\circ}\lesssim
\theta\lesssim 41^{\circ}$. For given angles $\theta$ the ratio of
the spin-orbit and tensor splittings appears to be close to unity
$a/t=1.0\pm 0.05,$ as it takes place for the $\chi_b$ and $\chi_c$
mesons. The value of tensor splitting $t$ has been defined from
the mass difference, $M(2^+) -M(1^+)$, which is not affected by
the coupling to the decay channel.

Calculated masses of the $2^+$ and $1^+$ states  are in good agreement
with experiment  for all $D, D_s, B, B_s$ mesons.

For the $0^+, 1^{+'}$ states, taking into account the mass shifts,
the following masses are predicted:

\begin{itemize}

\item $M(B^*_s) =5695(10)$ MeV which coincides with  $M(B^*)
=5695(10)$ MeV,

\item $M(B_{s}(1^{+'})) =5730(10)$ MeV,
 close to $M(B_{1}(1^{+'})) =5732(10)$ MeV.

\end{itemize}

\begin{acknowledgments}
The authors would like to thank for the support the President
Grant for scientific schools \# 843.2006.2. One of the authors
(M.A.T.) thanks RFBR for the partial support via grant \#
06-02-17120.
\end{acknowledgments}

\appendix

\section{Connection between the chiral quark-pion Lagrangian and the effective chiral Lagrangian}

The interaction of pions with quarks was introduced and developed
in \cite{21I}, \cite{22I}, see \cite{17I} for recent applications.
The effective chiral Lagrangian $\Delta L_{eff}$ contains one new
parameter $g^q_A$, \begin{equation}L_{\pi q} =\frac{g^q_A}{2f_\pi}
\bar \Psi\gamma_\mu \gamma_5 \vetau \Psi \partial^\mu
\vephi_\pi\label{a.1}\end{equation}and has the form of
pseudovector coupling, known from phenomenological applications in
the pion-nucleon systems. As was argued in \cite{21I},  $g^q_A$ at
large $N_c$ tends to unity.

In Eq.(17) we have used another form of the quark-pion
interaction, derived directly from the QCD Lagrangian in
\cite{20I} and not containing new
parameters,\begin{equation}\Delta L_{FCM}^{(1)}=\int \bar \psi (x)
\sigma |\vex| \gamma_5 \frac{\pi^a\lambda^a}{f_\pi} \psi(x) d^4
x.\label{a.2}\end{equation}In \cite{20I} the connection between
(\ref{a.1}) and (\ref{a.2}) was established and here we repeat the
derivation for the convenience of readers.

Consider  application of (\ref{a.2}) to the case of  pionic
transition  between states $\psi_1(x)$ and $\psi_2(x)$ of the
quark in heavy-light meson. Dirac equations for $\psi_i(x)$ can be
written as \begin{equation}(\veal\vep + \beta (m+\sigma |\vex|) +
V_{coul}) \psi_1 =\varepsilon_1 \psi_1\label{a.3}
\end{equation}\begin{equation}\bar \psi_2
(-\veal\vep+\beta(m+\sigma |\vex|)+V_{coul}) =\varepsilon_2\psi_2.
\label{a.4}\end{equation}Expressing in (\ref{a.3}), (\ref{a.4})
the term $\sigma|\vex|\gamma_5$ via $\veal\vep,\beta m$ etc. and
summing two equations, one gets
\begin{equation}\Delta L_{FCM}^{(1)} =\frac{1}{2 f_\pi}\bar \psi_2
(-2m\gamma_5\hat \pi +\beta \gamma_5 (\varepsilon_2-\varepsilon_1)
\hat \pi +\gamma_5 \beta \veal \vep \hat
\pi)\psi_1.\label{a.5}\end{equation}Since
$\gamma_i=-i\beta\alpha_i$, and $(\varepsilon_2-\varepsilon_1)
 \hat \pi = i \frac{\partial}{\partial t} \hat\pi (t)$ $\hat \pi
 (t) \sim e^{-i(\varepsilon_2 -\varepsilon_1) t}$, one can rewrite
 the last two terms in (\ref{a.5}) as $\gamma_\mu\gamma_5
 \partial_\mu\hat \pi$, and finally one arrives at
 \begin{equation} \Delta L_{FCM}^{(1)} =\frac{1}{2 f_\pi}\bar \psi_2 (-2 m\gamma_5 \hat
 \pi + \gamma_\mu\gamma_5 \partial_\mu \hat
 \pi)\psi_1.\label{a.6}\end{equation}
 Comparing (\ref{a.1}) and (\ref{a.6}), one can see that in the
 chiral limit, $ m_q\to 0$, two expressions coincide. However, for
 nonzero $m$, e.g. for strange quark  having the mass $m_s\sim
 0.2$ GeV at low scale $\sim 1$ GeV \cite{37}, first term in
 (\ref{a.6}) is becoming essential. Moreover, our expression
 (\ref{a.2}) is only the first term  in the   expansion of the
 exponent  (15) in powers of the pion field, and therefore this
  general Lagrangian can be  used for  decay channels with the production of two or several pions.

\section{Masses of heavy-light mesons}

To calculate masses and different matrix elements (m.e.) of a HL
meson ($q\bar b, q\bar c,$ or $\bar q b$) we use here the
relativistic string  Hamiltonian $\hat H_\omega$, derived in
\cite{24I}. For this Hamiltonian  the spin-averaged mass
$M_{cog}(nL)$   is given by simple formula:
 \begin{equation}M_{cog} (nL) = M_0 (nL)
+ \Delta_{SE} +\Delta_{str},\label{A.1}\end{equation}where $M_0$
is the eigenvalue (e.v.) of the spin-independent part $H_0$ of the
Hamiltonian $\hat H_{\omega}$, which  coincides  with well-known
spinless Salpeter Hamiltonian (SSH):
 \begin{equation}H_0 =\sqrt{\vep^2+m^2_q} +
\sqrt{\vep^2+m^2_b} + V_0(r),\label{A.2}\end{equation}
 \begin{equation}H_0\varphi_{nL} (r) =M_0\varphi_{nl}(r).\label{A.3} \end{equation}
 However, the mass (\ref{A.1}) contains negative (string)
correction and therefore in our approach
for a given  static potential  the levels
  with $L\neq0$ lie lower than for SSH. Also the mass (\ref{A.1})
 does not contain an overall fitting constant but takes into account
 NP self-energy term  $\Delta_{SE}$ for a light quark ( which is
calculated explicitly in \cite{38}).

 The static potential
 \begin{equation} V_0(r) =\sigma r -\frac43 \frac{\alpha_B(r)}{r}\label{A.4}\end{equation}is
 taken here from \cite{25I} with the vector oupling
 $\alpha_B(r)$ for $n_f=3$. The solutions of (\ref{A.3}) define
 $M_0(nL)$ and m.e., in particular,
 \begin{equation} \omega_q(nL) = \lan \sqrt{\vep^2+ m^2_q}\ran_{nL},~~ \omega_Q (nL)
 = \lan \sqrt{\vep^2+m^2_Q}\ran_{nL}, \label{A.5}\end{equation} which apear to be
 the dynamical (constituent) mass  of a light
 quark $\omega_q$ and $\omega_Q$ for a heavy    quark. Their values
 for  the $B$ and $B_s$ mesons are given in Table \ref{alla_table_7} together with  the
 reduced mass: $\omega_{red}
 =\frac{\omega_q\omega_b}{\omega_q+\omega_b}$.

\begin{table}
\caption{The constituent masses $\omega_q$ and $\omega_b$
 (in MeV) for the $B(1P)$ and $B_s(1P)$ mesons ($m_{u(d)} =0$, $m_s=200$
 MeV, $m_b=4780$ MeV)}
 \label{alla_table_7}
\begin{center}
\begin{tabular}{|c|c|c|}\hline
&&\\
& $B(1P)$ meson& $B_s(1P)$ meson\\
&&\\ \hline &&\\
$\omega_q(1P)$ &680&730\\&&\\ \hline &&\\
$\omega_b(1P)$ &4836&4840\\
&&\\\hline&&\\
$\omega_{red}$ &598&634\\&&\\\hline
\end{tabular}
\end{center}
\end{table}
As seen from Table \ref{alla_table_7}  the kinetic energy of a
light (strange) quark $\omega_q(1P)$ are not small and this fact
is important for the fine structure analysis.

 In the mass formula (\ref{A.1}) the correction
$\Delta_{SE}$ comes from NP self-energy contribution (which is
equal zero for the  $b$ quark), taken here in the simplest form
when self-energy contribution of the $c$ quark ($\lesssim -15$
MeV) can be neglected,because it is small  as compared to  the
pole $c$-quark mass known at present with the accuracy $\pm 100$
MeV \cite{26I}). For a light quark $\Delta_{SE}$ has been defined
in \cite{38}: \begin{equation}\Delta_{SE} (nL)=\delta - \frac{1.5
\sigma \eta_q}{\pi\omega_q},\label{A.6}\end{equation}in which
small correction,
 $\sim 3\div 6$ MeV (defined in \cite{25I}) is neglected.
The factor $\eta_{u(d)}=1.0$ for a light quark  and $\eta_s=0.65$
for the  $s$ quark with $m_s= 220 $ MeV.

The string correction  $\Delta_{str}$   for the $1P$ -wave
 $B(B_s)$ mesons is equal
$\Delta_{str}\approx -27(-21)$ MeV \cite{25I}. This negative
contribution to $M_{cog}$ improves an agreement with the
experimental  masses of $B(2^+)$ and $B(1^+)$ mesons \cite{28I}.
 In Table \ref{alla_table_8} the eigenvalues $M_0(1P)$ and $M_{cog}(1P)$
together with $\Delta_{SE}(1P)$ and $\Delta_{str} (1P)$ are given.


\begin{table}
\caption{The masses  $M_0, M_{cog}(1P)$  and   $\Delta_{SE}(1P),
\Delta_{str} (1P)$ (in GeV)  for the $B, B_s$ mesons  ($m_s=220$
MeV, $m_{u(d)}=0$, $m_b=4780$ MeV)} \label{alla_table_8}
\begin{center}
\begin{tabular}{|c|c|c|}\hline
&&\\
& $B(1P)$ & $B_s(1P)$\\\hline
&&\\ $M_0(1P)$& 5885&5925\\&&\\\hline &&\\

$\Delta_{SE}$&-126&-70\\ &&\\ \hline  &&\\

$\Delta_{str}$&-27&-20\\ && \\\hline  &&\\

$ M_{cog}(1P)$& 5.732&5.835\\&&\\\hline
\end{tabular}
\end{center}
\end{table}

The difference   between the e.v. $M_0(B)$ and $M_0(B_s)$ is only
$\sim 50$ MeV for $m_s=200$MeV, so that additional 50 MeV
difference  in $M_{cog} (1P)$ for the $B$ and $B_s$ comes from the
self-energy terms.

\section{Fine structure splittings}

To define  FS splittings of a HL meson we follow here the
approach, where the $\veL\veS$ basis is used  and
 the analysis of FS can be done in general terms \cite{29I}.
If one introduces tensor splittings $t(nP)$ and spin-orbit
splitting $a(nP)$ then
 the masses of the $2^+$ state $(j=\frac32)$ and $0^+$ state
 ($j=\frac12)$
can be written as  \begin{equation}M(2^+) = M_{cog} +a-0.1
t,\label{B.1}\end{equation}\begin{equation} M(0^+)
=M_{cog}-2a-t,\label{B.2}\end{equation}while
  the $1^+$ states, $~^3P_1$ and $~^1P_1$,  are   mixed.

The mixing matrix can be expressed  through the splittings $a$ and
$t$:

\begin{equation}\hat O_{mix} =\left( \begin{array}{ll} a-\frac76 t,&
-\frac{\sqrt{2}}{6} t\\-\frac{\sqrt{2}}{6} t,& -2a + \frac53
t\end{array}\right).\label{B.3}\end{equation}Then the  eigenvalues
and eigenvectors of this matrix define  ``higher'' and ``lower''
masses $M_H$ and  $M_L$ with $J^P=1^+$ and the decomposition of
their w.f.. The mass splittings are \begin{equation}M_H=M_{cog}
-\frac14 (2a-{t}) +\frac14 \sqrt{
(2a-{t})^2+32(a-t)^2};\label{B.4}\end{equation}\begin{equation}M_L=M_{cog}
-\frac14 (2a-{t}) -\frac14\sqrt{
(2a-{t})^2+32(a-t)^2}.\label{B.5}\end{equation}Each of these
levels is a decomposition of the  $~^3P_1$ and $~^1P_1$ states.
From (\ref{B.3}) it is evident that the weights in those
decompositions depend only on the ratio
\begin{equation}R=\frac{a}{t}.\label{B.6}\end{equation}Just the value of
this ratio defines the order of levels inside the $nP$ multiplet
and for given  $a/t$ the mixing angle $\theta$ in (9) can be
easily calculated. With the use of the relation (10) the
connection between the mixing angle $\phi$ in the basis
$|j=\frac12\ran, ~~ |j=\frac32\ran$ (see the definition in (6),
(7)) and the mixing angle $\theta$ in the $\veL\veS$ scheme, for
which
\[|1^+_H\ran =\cos\theta|~^3P_1\ran -\sin \theta
|~^1P_1\ran,\] have been  established: $\phi=-\theta
+35.264^{\circ}$.

The value  $a/t=1.0 \pm 0.05$ provides  large mass shifts of the
$D_s( 1^{+}_H)$ and  $B_s( 1^{+}_H)$ levels and at the same time
keeps the position of the $D_s( 1^+_L)$ and $B_s( 1^+_L)$
unchanged (with accuracy 2 MeV). Notice, that for all multiplets
with $a/t=0.95$  the mass difference between two narrow levels is
\begin{equation}M(2^+)-M(1^+_L)=1.31 t  {\rm ~for}~ a/t =
0.95,\label{C8a}\end{equation}\[ M(2^+)-M(1^+_L)=1.40 t  {\rm
~for}~ a/t = 1.0,\] so that this relation can be used to define
the parameter $t$ from experiment.

To interpret   this splitting $t$ one can use well-known
perturbative expression, taking one-gluon-exchange interaction and
neglecting higher in $\alpha_s$ corrections, which however may be
important \cite{31I}:
\begin{equation}t(nP) =\frac43 \frac{\alpha_{FS}\lan
r^{-3}\ran_{nP}}{\omega_q\omega_Q}.\label{B.7}\end{equation} Then
for $\alpha_{FS}=0.39$ one obtains the values: $t=11.7$ MeV for
the $B_s$ mesons $(\lan r^{-3}\ran_{1P} =0.080$ GeV$^3$) and
$t=12.1$ for the $B(1P)$ mesons $(\lan r^{-3}\ran_{1P} =0.0765$,
which are close to those, used in our analysis. For the $D(1P)$
and $D_s(1P)$ meson the   same $\alpha_{FS}=0.39$ gives
$t_D=t_{D_s}=29$ MeV $(\lan r^{-3}\ran_D=0.052$ and $\lan
r^{-3}\ran_{D_s}=0.055$ GeV$^3$).

However, in OGE approximation the situation with SO splitting is not so
simple.This
splitting can be presented in the convenient form from \cite{29I}
(here we also  keep the term proportional $m_Q^{-2}$:
 \begin{equation}a(nP)
=\left(\frac{1}{4\omega^2_q}+ \frac{1}{4\omega^2_Q}\right) A(nP) +
t(nP),\label{B.8}\end{equation}where the factor \begin{equation}
A(nP) =\frac43 \alpha_{FS} \lan r^{-3} \ran_{nP}- \sigma
 \lan r^{-1}\ran_{nP}.\label{B.9}\end{equation}To satisfy the condition
$a=0.95 t$ the term $A(1P)$ has to be small as compared with $t$.
However, in OGE approximation and for linear confinement this term
in (\ref{B.9}) appears to be negative and not small for
$\alpha_{FS}\sim 0.39$. The reason for that  needs a special
investigation and can mean that either higher radiative
corrections are important \cite{31I}, or a suppression of
confining potential in spin-orbit term ,observed on the lattice
\cite{32}, is essential.

 Notice that the Coulomb-type order of levels,  i.e. $M(0^+)<
 M(1^+_L)<M(1^+_H)< M(2^+)$, takes place only for not small
 ratio  $\frac{a}{t}\geq 0.606$. In our
 case with  $\frac{a}{t}=0.95 $  this condition is satisfied
  and the level $1^+_H$ lies below  the
  $2^+$ level,
 \begin{equation} M_H(1^+)<M(2^+).\label{B.11}\end{equation}Just this order of levels   is
 observed in the $D(1P)$ multiplet where the central value of the
 wide $1^{+'}$ level is smaller than the mass of $2^+$ state \cite{26I}.
 The FS splittings  of the $P$-wave HL mesons with $a/t=0.95$ are given in
 Tables \ref{alla_table_2}, \ref{alla_table_3}, \ref{alla_table_5}.

\end{document}